\documentclass [aps,preprint,showpacs]{revtex4}
\usepackage{graphicx}
\begin{document}
\title{Many-body correlations and Isospin equilibration in multi-fragmentation processes}
\author{M.Papa$^{a)}$\footnote{e-mail: papa@ct.infn.it} and G.Giuliani $^{b)}$}
\affiliation{\textit{ a) Istituto Nazionale Fisica Nucleare-Sezione
di Catania, V. S.Sofia 64 95123 Catania Italy}}
\affiliation{\textit{
 b)Dipartimento di Fisica e Astronomia, Universit\'a di Catania
 V. S.Sofia 64 95123 Catania Italy}}
\begin{abstract}
Isospin equilibration in multi-fragmentation processes is studied
for the system $^{40}Cl+^{28}Si$ at 40 MeV/nucleon. The
investigation is performed through semiclassical microscopic
many-body calculations based on the CoMD-II model. The study has
been developed to describe isospin equilibration processes involving
the gas and liquid "phases" of the total system formed in the
collision processes.
 The investigation of the behavior of this observable in terms of the
 repulsive/attractive action of the symmetry term, highlights
 many-body correlations which are absent in semiclassical mean-field
approaches.
\end{abstract} 
\maketitle

An interesting subject related to the Heavy Ions Isospin physics
\cite{schr} is the process  leading to the equilibration of the
charge/mass ratio between the main partners of the reaction. The so
called "isospin diffusion" phenomenon has been indicated as the
associated mechanism acting between the reaction partners
\cite{pawel,betty,baodif,dtoro}. In particular, in basically binary
processes occurring after the collision of $Sn$ 124 and 112 isotopes
\cite{betty}, evidence of partial equilibrium in the charge/mass
ratios of the quasi-projectile (QP) and quasi-target (QT) has been
deduced through the study of the iso-scaling parameters related to
the isotopic distributions. In this case dynamical calculations
based on the BUU model \cite{buu} show that the degree of
equilibration depends on the behavior of the symmetry potential
$U^{\tau}$ as a function of the density. The analysis in this kind
of studies, however, is based on the linear relation between the
iso-scaling parameter and the relative neutron excess $Y$ of the
emitting sources (an assumption typical of several statistical
models), moreover, it is also assumed that both quantities  weakly
depend on the secondary statistical decay processes. However, in
general, this last condition depends on the fragment excitation
energies, or temperatures, and on the distinctive features  of the
 models used to simulate the second stage decay \cite{betty1}.
In this work we want to describe the isospin equilibration process,
looking at the whole system, by using the dynamical variable
$\overrightarrow{V}(t)=\sum_{i=1}^{Z_{tot}}\overrightarrow{v}_{i}$.
The sum on the index $i$ is performed  on all the $Z_{tot}$ protons
of the system. $\overrightarrow{V}(t)$  corresponds to the time
derivative of the total dipole of the system. The velocities
$\overrightarrow{v_{i}}$ are computed in the center of mass (c.m.)
reference system. Several studies were based on this dynamical
variable to describe pre-equilibrium $\gamma$-ray emission (see
Refs.\cite{asygdr,ca10mev,trasth,trasex1} and references therein ).
Various reasons  dictate this choice to describe also isospin
equilibration processes.

 - i) After the pre-equilibrium
 stage at the time $t_{pre}$, when a second stage characterized by
 an average isotropic emission of the secondary sources (statistical equilibrium)
 takes place, the ensemble average satisfies the following relation:
 $\overline{\overrightarrow{V}}(t_{pre})=
 \overline{\overrightarrow{V}}(t>t_{pre})\equiv\overline{\overrightarrow{V}}$ \cite{ca10mev}.
The average value of the dynamical variable at $t_{pre}$
 is in fact invariant with respect to statistical processes
 and therefore $\overline{\overrightarrow{V}}$ is an interesting
 observable to be investigated.
 In particular, $\overline{\overrightarrow{V}}$
 can be expressed as function of the charge $Z$, mass $A$, average multiplicity
 $\overline{m}_{Z,A}$ and the mean momentum
 $\langle \overrightarrow{P}\rangle_{Z,A}$ of the detected particles having charge $Z$
 and mass A in the generic event:
\begin{eqnarray}
\overline{\overrightarrow{V}}=\sum_{Z,A}\frac{Z}{A}\overline{m_{Z,A}}
\overline{\langle \overrightarrow{P} \rangle}_{Z,A}
C_{\langle \overrightarrow{P}\rangle}^{Z,A}\\
C_{\langle \overrightarrow{P}\rangle}^{Z,A}=
\frac{ \overline{m_{Z,A}\langle \overrightarrow{P}\rangle_{Z,A}}}
{\overline{\langle \overrightarrow{P}\rangle}_{Z,A}\overline{m}_{Z,A}}
\end{eqnarray}
$C_{\langle \overrightarrow{P}\rangle}^{Z,A}$ is the correlation
function between the multiplicity and the mean momentum. This
correlation function plays a determinant role for the invariance
property and therefore asks for an analysis event by event in which
many-body correlations can not be neglected.

$\overline{\overrightarrow{V}}$, for symmetry reasons, lie on the
reaction plane. As suggested from Eq.(1), it is directly linked with
a  weighted mean of the charge/mass ratio. It takes into account
also the average isospin flow direction through the momenta
$\overline{\langle \overrightarrow{P}\rangle}_{Z,A}$.

 -ii) In the general case, we
find attractive the following decomposition:
$\overline{\overrightarrow{V}} = \overline{\overrightarrow{V}_{G}}+
\overline{\overrightarrow{V}_{L}}+\overline{\overrightarrow{V}_{GL}}$
where with $\overline{\overrightarrow{V}_{G}}$ and
$\overline{\overrightarrow{V}_{L}}$ we indicate the dipolar signals
associated to the gas "phase" (light charged particles) and to the
"liquid" part, corresponding to the motion of the ensemble of the
produced fragments (QP,QT, if any, and intermediate mass fragments).
The signal $\overline{\overrightarrow{V}_{GL}}$ is instead
associated to the relative motion of the two "phases". By supposing,
for simplicity, that the gas "phase" is formed by neutrons and
protons, $\overline{\overrightarrow{V}}$ can be further decomposed
as:
\begin{eqnarray}
\overline{\overrightarrow{V}}=
 \overline{\frac{A_{G}(1-Y^{2}_{G})}{4}\overrightarrow{v}_{r}^{NP}}+
 \overline{\frac{\mu_{G,L}(Y_{L}-Y_{G})}{2}\overrightarrow{v}_{cm,LG}}+
 \overline{\overrightarrow{V}_{r,L}}
\end{eqnarray}
In the above expression the first term represents the contribution
related to the neutron-proton relative motion of the gas "phase"
expressed through the relative velocity
$\overrightarrow{v}_{r}^{NP}$, the second term is related to the
relative motion $\overrightarrow{v}_{cm,LG}$ between the centers of
mass of the "liquid" complex and the "gas"; the last term represents
the contribution produced by the relative motion of the fragments. A
similar expression can be obtained including in the gas "phase"
other light particles. From this decomposition we can see how the
isospin equilibration condition ($\overline{\overrightarrow{V}}=0$),
for the total system, requires a very delicate balance which depends
on the average neutron excess of the produced "liquid drops"
$\overline{Y_{L}}$, on the one associated to the gas "phase"
$\overline{Y_{G}}$ and on the relative velocities
 between the different parts. To enlighten the role played by some of the terms reported in Eq.(3),
 we can discuss the idealized decay of a
charge/mass asymmetric source through  neutrons and protons emission
(or the case in which the liquid drops are produced through a
statistical mechanism $\overline{\overrightarrow{V}_{r,L}}=0$).
Moreover, for simplicity, we can consider uncorrelated fluctuations
between the velocities and neutron excesses. In absence of
pre-equilibrium emission or for identical colliding nuclei
($\overline{\overrightarrow{v}_{cm,LG}}= 0$) the second term of
Eq.(3) is zero, and the isospin equilibration will require a
proton-neutron symmetric gas "phase" and/or absence of relative
neutron-proton motion. For non identical colliding nuclei, if
pre-equilibrium emission is present, then
$\overline{\overrightarrow{v}_{cm,LG}}\neq 0$. In this case, if
$\overline{Y_{G}}\neq \overline{Y_{L}}$, as due to the isospin
"distillation" phenomenon, the first term has to be necessarily
different from zero and it will contribute to the neutron-proton
differential flow $F_{np}$ (see the following). Therefore, for the
whole system, the isospin equilibration process can not be explained
only in terms of isospin diffusion and drift between the main
partners of the reaction. It is in fact necessary to take into
account the gas "phase" contribution which we can regard as a kind
of "dissipation" with respect to the system formed by the liquid
part. In particular in this work, as an example, we will discuss the
results obtained through the Constrained Molecular Dynamics-II
approach (CoMD-II) \cite{comd1,comdII} applied to the charge/mass
asymmetric system $^{40}Cl+^{28}Si$ at 40 MeV/nucleon. The study is
performed by using different options for the symmetry potential term
$U^{\tau}$. Nevertheless, to clearly understand the dynamics of the
investigated system, it is necessary to dedicate the following part
to a discussion of this term. In particular, we will illustrate that
the existence of many-body correlations, as produced in our
approach, strongly affects this part of the interaction. \vskip 7pt
In the framework of the present version of the CoMD-II model the
isospin momentum dependent part of the total energy is included
through the Pauli principle constraint and the potential symmetry
term has been implemented as follows \cite{asygdr,comdII}:
\begin{eqnarray}
  U^{\tau} &=& \frac{a_{sym}}{2s_{g.s.}}F'(s)\beta_{M} \\
  \beta_{M}  &=& \rho^{NN}+\rho^{PP}-\rho^{NP}-\rho^{PN}\\
  \rho^{KK'} &=& \sum_{i\neq j}^{i\subseteq K,j\subset K}\rho_{i,j} \;\;\;\;\;\;\;\;\;\; K,K'=N\;\; or\;\; P\\
  F'(s)&=& \frac{2s}{s_{g.s.}+s} \;\;\;\;\;\;\;\;\;\; Stiff1\\
  F'(s)&=& 1\;\;\;\;\;\;\;\;\;\;\;\;\;\;\; \;\;\;\;\;Stiff2 \\
  F'(s) &=&(\frac{s}{s_{g.s.}})^{-\frac{1}{2}}\;\;\;\;\;\;\;\; Soft \\
  s &=&\frac{4}{3A} \sum_{i \neq j}^{i,j\subseteq K\cup K'}\rho_{i,j}
\end{eqnarray}
With the superscript N and P we indicate the neutron and the proton
ensemble of nucleons respectively and A is the total mass of the
system.
 $\rho_{i,j}$ represents the normalized Gaussian overlap
matrix elements \cite{comd1} typical of the quantum molecular
dynamics approach \cite{aich}. The label $g.s.$ indicates the ground
state configuration. The positive form factors $F'(s)$ are such that
$F'(s)\frac{s}{s_{g.s.}}$ has the same functional form like $F(u)$
proposed in \cite{bao0} (see Eq.(3) in Ref.\cite{bao0}). They have
been used extensively in mean-field (M.F.) approaches. Apart from a
normalization factor, equal to $\frac{4}{3}$, $s$ is related to the
average (with respect to the number of two-body interactions)
overlap integral . For the system under study  and for compact
configurations $s$ is well approximated by the average one-body
density $\rho$. The factor $\beta_{M}$ instead arises naturally from
the many-body approach. It takes into account explicitly, through
the last two negative terms in Eq.(5), that the microscopic two-body
nucleon-nucleon interaction in the isospin singlet states is more
attractive than the one related to triplet states \cite{asygdr}
($a_{sym}=72 MeV$).  For the moderately asymmetric system, here
investigated, self-consistent calculations, including the ones for
the searching of the g.s. properties, produce a negative value of
$\beta_{M}$. This result is due to the combined actions of the more
attractive force, for the isospin singlet states, and the repulsive
action of the Coulomb interaction for proton-proton couples. We have
verified also that the Pauli principle plays a role even if with a
smaller extent.
 These correlations, in fact, tend to increase the average (with respect the number
 of two-body interactions) neutron-proton overlap
integral $\tilde{\rho}^{NP}$ and to decrease the overlap
$\tilde{\rho}^{PP}$ and $\tilde{\rho}^{NN}$ related to the
neutron-neutron and proton-proton couples. These average overlap
integrals are defined as: $\tilde{\rho}^{II'}=\rho^{II'}/II'$ with
$I$ and $I'$ equal to $N$ and/or $P$.
 In our calculations we have verified that $\tilde{\rho}^{NP}=\tilde{\rho}^{PN}\equiv\alpha\frac{(\tilde{\rho}^{PP}+
 \tilde{\rho}^{NN})}{2}$
with $\alpha\simeq 1.1-1.2$. Therefore, at low asymmetry, it is
enough a small correlation effect to produce a negative value of
$\beta_{M}$, as due to the structure of Eq.(5).
 However, it results that
$\beta_{M}$ decreases by increasing the symmetry of the system (the
number of singlet states increases) and, on the contrary, it
increases by decreasing the symmetry of the system. These changes
follow a parabolic dependence with respect the $\chi=N-Z$ variable.
 This means that $\beta_{M}$ contains, independently from the sign,
 the right behavior necessary to explain the differences in the
binding energies of isobars nuclei and it is able to generate the so
called "isospin distillation" phenomenon \cite{dtoro} in the
dynamics of hot sources.
 However, the negative sign of
$\beta_{M}$ has  relevant consequences in to determine the effects
of the different options for $U^{\tau}$ on the dynamics of the
investigated processes. For example, contrary to the semi-classical
M.F. description, the so called Stiff cases (in spite of this change
of sign, we retain the same nomenclature to indicate the different
options) show a decreasing behavior with the density.
 The Soft case, here investigated, even if negative,
maintains the increasing pattern typical of a M.F. approach (see
Fig.5(c)). \noindent We conclude this part by briefly describing how
in M.F. and/or in Liquid Drop Model  these correlations are washed
out. For this purpose we assume, for simplicity, $A,N,Z\gg 1$, and
we rewrite $\beta_{M}$ as: $\beta_{M}=N^{2}\tilde{\rho}^{NN}+
Z^{2}\tilde{\rho}^{PP}-2NZ\tilde{\rho}^{NP}$. \textit{M.F.
approximation is now easily obtained by supposing
$\tilde{\rho}^{II'}=\frac{\sum_{i\neq j=1}^{A}\rho_{i,j}}{A^{2}}
\equiv\tilde{\rho}=\frac{3}{4A}\rho$ to be independent on $I$ and
$I'$}. This means that the correlation producing the aforementioned
differences in the matrix elements $\tilde{\rho}^{II'}$  are
averaged through the overlap integral $\tilde{\rho}$, so that :
$\beta_{M}\rightarrow\frac{3}{4}\frac{\rho}{A}(N-Z)^{2}=\frac{3}{4}\rho^{2}V
\frac{(\rho_{n-\rho_{p}})^{2}}{\rho^{2}}$, where $\rho_{n}$,
$\rho_{p}$ and $\rho$ represent the neutron, proton and total
densities respectively. V is the volume. In this case, therefore, we
obtain that the interaction depends locally only through one-body
densities and it is positive defined. Taking into account  that for
the present system, in compact configurations, $s\cong\rho$, by
substituting the right end of the above relation in Eq.(4) and
dividing by the volume $V$, we get a precise correspondence with the
symmetry energy density used in Ref.\cite{bao0} if $e_{0}=27 MeV$.
This value includes the reduction of the Fermi motion as due to
finite size effects and corresponds to an $S_{0}$ value  (see Eq.(4)
of Ref.\cite{bao0}) of about 36 MeV. Finally we note that these
correlations not only change the sign of the symmetry energy, but
they heavily affect also the strength. In particular, according to
Eqs.(4-10) and to the definition of $\beta_{M}$ in terms of the
average overlap integral $\tilde{\rho}^{II'}$  it is straightforward
to evaluate the contribution of $\beta_{M}$ by assuming  10\% of
correlation, i.e. $\alpha=1.1$. By considering a total system like
the one investigated in the present work (A=68 Y=0.088) at normal
density we obtain $\beta_{M}\simeq -0.43 fm^{-3}$ in the correlated
case and $\beta_{M}=0.065 fm^{-3}$ in the M.F case. According to
this evaluation we observe that also with a correlation value of the
order of 1\% the induced effect on the symmetry term is not
negligible. This strongly suggests that the qualitative effects
discussed in the present work can be still present by using other
kinds of effective interactions. We have to note also that,
according to the increasing behavior of the symmetry energy with the
charge/mass asymmetry, for systems with  mass around 64 units, the
$\beta_{M}$ factor becomes positive for relevant asymmetry
$|Y_{c}|>0.25$ restoring the repulsive behavior of the asy-stiff
cases here investigated.\vskip 7pt

Now we can discuss the results concerning the isospin equilibration
process for the system $^{40}Cl+^{28}Si$ at 40 MeV/nucleon. For the
system under study, in Fig. 1(a) and Fig. 1(b) we show the average
total dipolar signals as evaluated through CoMD-II calculations
along the $\hat{z}$ beam direction $\overline{V}^{z}$ and along the
impact parameter direction $\hat{x}$, $\overline{V}^{x}$,
respectively. The reference frame is the c.m. one. The impact
parameter $b$ is equal to 4 fm.
\begin{figure}
\includegraphics[height=9cm, width=9cm, angle=0]{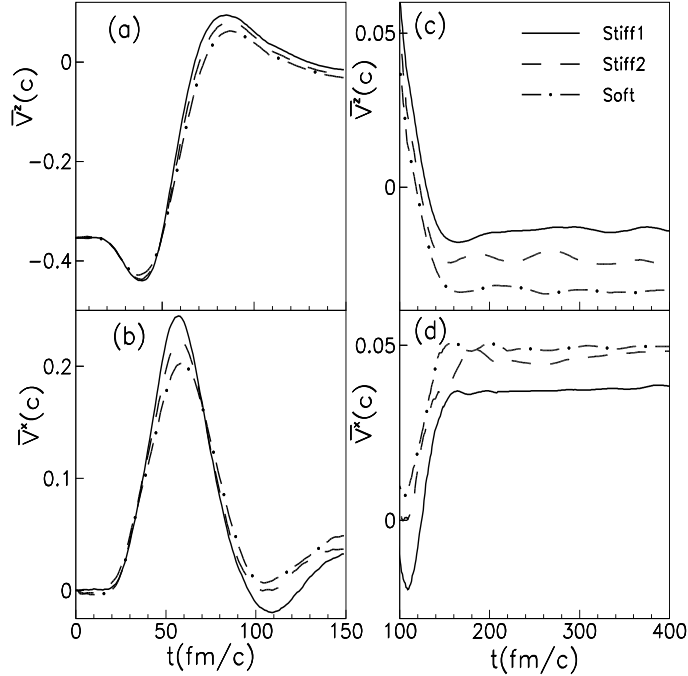}\\
\caption{Average dipolar signals along the $\hat{z}$ and $\hat{x}$
directions are plotted as a function of time in the intervals  0-150
fm/$c$ (panels (a) and (b)) and 100-400 fm/$c$. The three kinds of
lines refer to the different options (see eqs.(7-8)) describing the
symmetry potential.}
\end{figure}

In these figures the average dipolar signals are shown for the first
150 fm/$c$. Different lines refer to different symmetry potentials.
The isospin independent compressibility is equal to 220 MeV
according to Ref. \cite{comd1}. In the first 150 fm/$c$ we can see
that in all the cases wide oscillations are present. They are
responsible for the pre-equilibrium $\gamma$-rays emission
\cite{asygdr,ca10mev}. The damped oscillations tend to smaller and
constant values (within the uncertainty of the statistics of the
ensemble average procedure) as can be seen in Fig. 1(c) and Fig.
1(d) in which the dynamical evolution is followed from 100 fm/$c$ up
to 400 fm/c. The time interval in which the stationary behavior is
reached corresponds to the average time for the formation of the
main fragments. In Fig. 2(a), for the different symmetry potentials,
we plot the charge $Z$ distributions. The Stiff1 and Stiff2 options
clearly show a multi-fragmentation pattern, while for the Soft case
the behavior is closer to an exponential trend making reliable a
"vaporization" scenario.
\begin{figure}
  \includegraphics[height=9.cm, width=9.cm, angle=0, keepaspectratio]{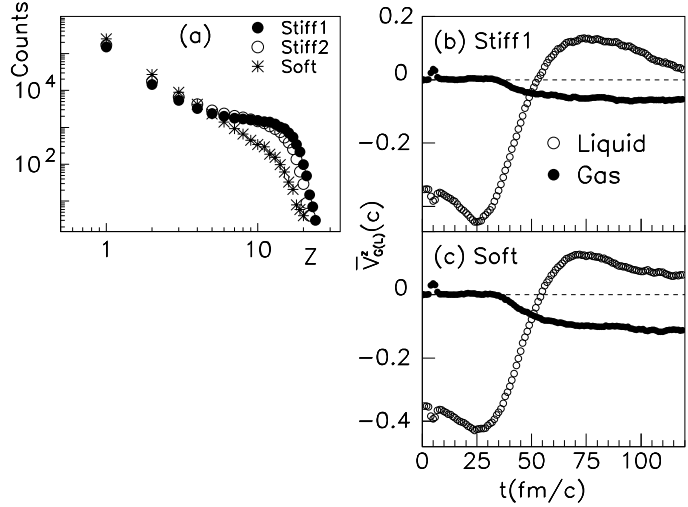}\\
  \caption{ In the panel (a) we show the charge $Z$ distributions of
the reaction products for the different options describing the
symmetry potential (see eqs.(7-8)). In panels (b) and (c), for the
Stiff1 and Soft cases, the $\hat{z}$ components of the dipolar
signals are shown for the liquid and gas "phases".}
\end{figure}

 We now briefly comment the results already
shown in Figs. 1(c) and 1(d). We can see that the asymptotic values
sensitively depend on the options used for the symmetry term. For
example, concerning the dipolar signal along the $\hat{z}$
direction, the difference between the Stiff1  and the Soft cases is
about 85\% while  between the Stiff1 and the Stiff2 options is about
53\%. The changes of the dipolar signals along  the $\hat{x}$
direction are smaller of the order of 10-30\%. To understand the
role of the "gas" particles, in Fig. 2(b) and in Fig. 2(c) we show
with open circles and with closed circles the values of dipolar
signal along the $\hat{z}$ direction, $\overline{V}_{G}^{z}$ (gas
"phase") as a function of time and $\overline{V}_{L}^{z}$ (liquid
"phase"), for the Stiff1 and Soft parameterizations respectively. We
can see that the contributions related to the "gas" are negative
(light particles coming essentially from the target). They are quite
effective in to reduce the absolute value of the total dipolar
signals driving the total system through the isospin equilibration.
It is also possible to see that $\overline{V}_{G}^{z}$ shows a
larger absolute value for the Soft case with respect the Stiff1
option. This is due to the more repulsive effect obtained in the
Soft case with respect to the other ones. In Fig. 3(a) we show, as a
function of time, the average overlap integral $s$ for the three
cases.

\begin{figure}
  \includegraphics[height=9.cm, width=9.cm, angle=0, keepaspectratio]{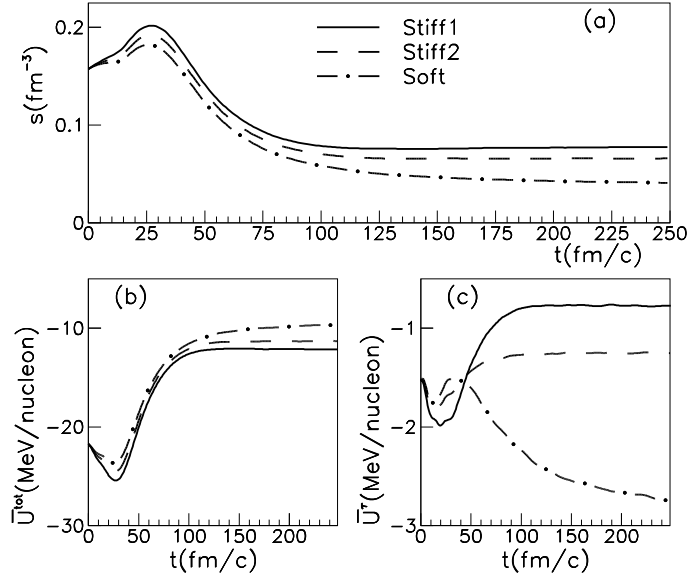}\\
  \caption{ (a) average overlap integral $s$, (b) average total
potential, (c) average symmetry potential are shown as a function of
time for the different options describing the symmetry potential
(see eqs.(7-8)). }
\end{figure}

In Fig. 3(b) and Fig. 3(c) we show the corresponding total average
potential $\overline{U}^{tot}$ and the symmetry one
$\overline{U}^{\tau}$. From Fig. 3(a) we can see that the maximum
value of $s$, $s_{max}$, is reached in about 30 fm/c. The Soft case
shows the lower $s_{max}$ value, the Stiff1 the higher one (about
1.25 times the $s_{g.s.}$ value). Accordingly, $\overline{U}^{tot}$
displays the less attractive behavior  for the Soft case (see Fig.
3(b)). The reasons of these differences can be understood by looking
at Fig. 3(c). $\overline{U}^{\tau}$ shows, in fact, a clear
repulsive behavior as a function of the density for the Soft case
while the Stiff1 case gives rise to an attractive effect. At later
times,  when the density is decreasing and the fragment formation
process takes place for $s<s_{g.s.}$, the behavior of
$\overline{U}^{\tau}$ in the different cases is reversed. The Stiff2
case shows always an attractive behavior with intermediate values
concerning the strength. The  differences observed around 30 fm/$c$
play a crucial role in to determine the later evolution of the
system \cite{lynch}. This can be understood by looking at the
figures 2(a) and 2(b). In these figures it is in fact clearly
visible that the dipolar signals associated to the gas "phase" start
to develop just at 30 fm/$c$. The less attractive behavior
$\overline{U}^{tot}$ for the Soft case determines a lower collision
rate and a higher multiplicity of light particles. For example, the
total number of neutrons and protons emitted at 160 fm/c is, on
average, about 15 for the Soft case and 11 for the Stiff1 one. The
gas "phase" is more rich in neutron for the Soft case
($Y_{G}\cong0.23$) with respect to the Stiff1 option
($Y_{G}\cong0.18$). These values are rather large as compared with
the neutron excess fraction of the total system and with the one
related to the biggest fragment. This clearly  indicates the
occurrence of the so called "isospin distillation". As above
mentioned the dipolar signals associated to the relative
neutron-proton motion for the gas "phase" can affect the
neutron-proton differential flow $F_{np}$ \cite{flowg}. As an
example in Fig. 4(a) for the Soft and Stiff1 cases we show the
neutron-proton differential flow as a function of the particles
rapidity $y$ normalized to the projectile one $y_{beam}$.

\begin{figure}
  \includegraphics[height=9.cm, width=9.cm, angle=0, keepaspectratio]{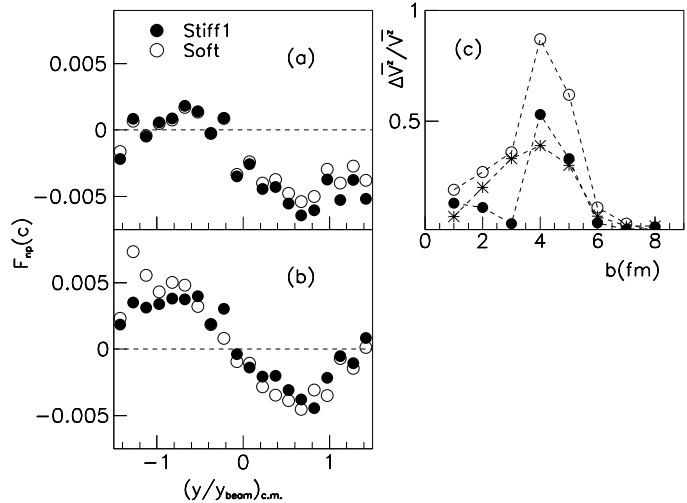}\\
  \caption{(a) Differential neutron-proton flow $F_{np}$ as a function
of $(\frac{y}{y_{beam}})_{c.m.}$ (see text). (b) The same quantity
is plotted after the corrections given by the relative
neutron-proton motion related to the gas "phase". (c) For different
symmetry potentials we show the relative changes
$\frac{\Delta\overline{V}^{z}}{\overline{V}^{z}}$ of the  average
total dipolar signal along the $\hat{z}$ component as a function of
the impact parameter.}
\end{figure}
The rapidity values are evaluated in the c.m. reference system. In
Fig. 4(b), we show the same quantities corrected for the average
free neutron-proton relative motion. We can see that the differences
are considerable and, in this case, they are due essentially to the
$\hat{x}$ component of the relative motion. We conclude this study
by showing in Fig. 4(c), for the different symmetry terms, the
relative changes $\frac{\Delta\overline{V}^{z}}{\overline{V}^{z}}$
of the average total dipolar signals along the $\hat{z}$ component
as function of the impact parameter. It is clearly visible the
higher sensitivity obtained for the impact parameter range $b\cong
3.5-5.5$ fm. CoMD-II calculations show that, for more central
collisions, even if the relative changes in the overlap integrals
are slightly more pronounced, the increasing of the collision rate
produces a more damped mechanism with respect to the dipolar degree
of freedom. This determines a smaller sensitivity to the change of
the symmetry energy. On the contrary, for more peripheral reactions,
the dipolar signal shows a smaller damping and a related smaller
degree of equilibration. In this case, however, the changes in the
relative overlap integral $s$ are smaller so that the associated
changes in $\overline{U}^{\tau}$ are less pronounced. These
suppression mechanisms and the related existence of an optimal range
of impact parameters for an enhanced visibility of the isospin
effects seem to us quite general, and clearly depending also on the
way in which the collision term is treated \cite{comdII}. \vskip 7pt

In summary, in this work the isospin equilibration process for the
asymmetric charge/mass system $^{40}Cl+^{28}Si$ at 40 MeV/nucleon
has been investigated by studying  the ensemble average of the time
derivative of the total dipole $\overline{\overrightarrow{V}}$ as
evaluated through CoMD-II calculations. Some general properties of
this quantity have been discussed. In particular, it allows to give
a consistent definition of isospin equilibration also in complex
reactions evolving through multi-fragmentation processes. CoMD-II
calculations show that the  asymptotic values of
$\overline{\overrightarrow{V}}$ for these processes are quite
sensitive to different symmetry potential options; moreover, for
this system, the dipolar contribution associated to the
pre-equilibrium charged particles emission is relevant in to
determine the value of $\overline{\overrightarrow{V}}$. Detailed
CoMD-II calculations also enlighten fundamental differences in the
behavior of the symmetry potential with respect to a description
based on a M.F. approach. The negative sign of the leading
$\beta_{M}$ factor in the symmetry potential  represents a typical
example in which the ensemble average, performed on many
realizations of a semiclassical many-body dynamics approach, gives
arise to deep differences with respect to the average behavior
evaluated with a M.F. approach. These deviations are due to the
highly correlated neutron-proton motion generated by the symmetry
term and they are clearly beyond the one-body description. Moreover,
the calculations presented in the present work suggest the existence
of relevant critical charge/mass asymmetries starting from which the
symmetry potential changes sign producing a transition from a
density attractive to a density repulsive behavior for the asy-stiff
cases. We conclude by observing that this subject needs further
detailed studies  in the next future together with a specific
investigation about the effects of these correlations on the
behavior of others isospin-dependent observables.

\vfill \eject
\end{document}